\documentclass[12pt]{article}
\usepackage[centertags]{amsmath}
\usepackage{graphicx}
\usepackage{amssymb,amsfonts}
\usepackage[comma,super]{natbib}
\usepackage{latexsym}
\begin{document}
\title{Shearing radiative collapse with expansion and acceleration}
\maketitle
%\vspace{1.5cm}
%{\bf
\begin{center}
\author {S. Thirukkanesh\footnote{Permanent address:
Department of Mathematics, Eastern University, Chenkalady, Sri
Lanka.\\
},
S.S. Rajah\footnote{Permanent address: Department of Mathematics, Durban University of Technology,
Steve Biko Campus,  Durban, 4001, South Africa.\\
} and S. D. Maharaj\footnote{Electronic mail: maharaj@ukzn.ac.za}\\
Astrophysics and Cosmology Research Unit,\\
School of Mathematical Sciences,
University of KwaZulu-Natal,\\
Private Bag X54001,
Durban, 4000,\\
South Africa.}\\
\end{center}

\begin{abstract}

We investigate the behaviour of a relativistic spherically symmetric
radiative star with an accelerating, expanding and shearing interior matter
distribution in the presence of anisotropic pressures. The junction condition can be written
in standard form in three cases: linear, Bernoulli and Riccati equations. We can integrate the boundary condition
in each case and three classes of new solutions are generated. For particular choices of the metric we
investigate the physical properties and consider the limiting behaviour for large values of time.
The causal temperature can also be found explicitly.
\end{abstract}

\section{Introduction}
The problem of radiative gravitational collapse was first
investigated by Oppenheimer and Snyder \cite{1}. Their interior
spacetime is represented by a Friedmann-like solution for an
isotropic homogeneous universe, and the exterior spacetime is
described by the exterior Schwarzschild metric. The process of gravitational
collapse is highly dissipative. Therefore heat flow in the interior of the star must be present, and taken into account
so that the interior solution of the radiating
star can match to the Vaidya \cite{2} exterior metric at the
boundary. The investigation of the gravitational behaviour of a
collapsing star depends on the determination of the junction
conditions matching the interior metric with the exterior Vaidya metric
across the boundary of the star. Santos \cite{3} formulated the
junction conditions for a shear-free fluid distribution with
isotropic pressures and made it possibile to complete the model. His treatment paved the way to investigate physical
features such as surface luminosity, dynamical stability, relaxation
effects and temperature profiles. Raychaudhuri \cite{4}, showed that
the slowest collapse arises in the case of shear-free fluid interiors.
Kolassis \emph{et al} \cite{5} assumed geodesic fluid trajectories
when generating an exact model. Their model was generalised to include several new classes of solution
in geodesic motion by Thirukkanesh and Maharaj \cite{6}. In the past many investigations in
radiating collapse have focussed on shear-free spacetimes with
isotropic pressures (see the treatments of Herrera \emph{et al}
\cite{7}, Maharaj and Govender \cite{8}, Herrera \emph{et al} \cite{9}
and Misthry \emph{et al} \cite{10}).

The next stage of development was to include shear in the model of a radiating star. Naidu \emph{et al} \cite{11} included
anisotropic pressures in the presence of shear for the interior spacetime and found simple exact solutions
for geodesic fluid trajectories. This toy model was generalised by Rajah and Maharaj \cite{12}
by demonstrating solutions to a Riccati boundary equation
governing the gravitational behaviour. The general situation requires a model
which is expanding, accelerating and shearing. Noguiera and Chan \cite{13}, modelling shear viscosity and bulk viscosity,
attempted such a study but found that they needed to utilise
numerical techniques to make progress. Some recent progress has been made in finding exact models for 
Euclidean stars by Herrera and Santos \cite{14} and
Govender \emph{et al} \cite{15}. In Euclidean stars with shear the areal radius and proper radius are equal throughout the evolution of the
radiating star. Our objective here is to show that it is possible to solve the relevant equations, for the general case,
 exactly in a systematic fashion.
 Our approach is the first analytic treatment to consider exact models with all the kinematical quantities present. We believe that these solutions will be helpful in studying physical features of a relativistic star in an astrophysical setting.

In this paper we attempt to perform a systematic treatment of the
governing equation at the boundary of the relativistic star with the interior consisting
of a fluid which has nonzero acceleration, expansion and shear. The junction condition is a
nonlinear partial differential equation containing all three metric functions of spherical symmetry.
In Section 2, we derive the field equations and the junction conditions. In Section 3, we give the boundary differential equation
governing the gravitational behaviour of a radiating, shearing and
accelerating sphere. In Section 4, three new classes of exact solutions to the boundary condition are found in closed form. In Section 5,
we briefly investigate the physical features of the model generated and present the explicit form of the causal temperature for a particular choice
of the metric functions. Some concluding remarks are made in Section 6.

\section{The model}
The most general form for the interior space time of a spherically
symmetric collapsing star, which is expanding, accelerating and shearing, is given by the line metric
\begin{equation}
\label{eq:g1} ds^2 =-A^2 dt^2 +B^2 dr^2 +Y^2(d\theta^2 +\sin^2
\theta d\phi^2),
\end{equation}
where $A, B$ and $Y$ are in general functions of both the temporal coordinate
$t$ and the radial coordinate $r$. The existence of a fluid 4-velocity vector
$\mathbf{u}$ enables us to introduce the kinematical
quantities
\begin{equation} {\dot u}^a = u^{a}{}_{;b}u^b, \,\,\,
\Theta = u^a{}_{;a}, \,\,\, 
\sigma_{ab} = h_a{}^c h_b{}^d u_{(c;d)},
\end{equation}
where $h_{ab} = g_{ab} + u_au_b$ ($h_{ab}u^a =0$) is the symmetric projection tensor. The acceleration vector ${\dot u}^a$
(${\dot u}^a u_a=0$) represents the acceleration of the fluid particles relative to the congruences of $\mathbf{u}$;  
the expansion scalar $\Theta$ measures the rate of increase of a volume of fluid element; 
the  shear $\sigma_{ab}$ ($\sigma_{ab}u^b =0 = \sigma^a{}_a$) represents the  tendency of a sphere to distort to an ellipsoid.
For the comoving fluid 4-veloctiy  $u^a =\frac{1}{A}{\delta}_0^a$ and  the line element (\ref{eq:g1}),
 the acceleration vector $\dot u^a$, the
expansion scalar $\Theta$ and the magnitude of the shear scalar
$\sigma$ are given by
\begin{subequations}
\begin{eqnarray}
\label{eq:g2a}\dot{u}^a &=& \left(0, \frac{A'}{AB^2},0,0\right),\\
 \label{eq:g2b}\Theta &=&
\frac{1}{A}\left(\frac{\dot{B}}{B}+2
\frac{\dot{Y}}{Y}\right),\\
\label{eq:g2c}\sigma &=& -\frac{1}{3
A}\left(\frac{\dot{B}}{B}-\frac{\dot{Y}}{Y}\right),
\end{eqnarray}
\end{subequations}
where primes and dots on the metric functions denote differentiation with respect to $r$
and $t$ respectively. The energy momentum tensor for the interior
matter distribution has the form
\begin{equation}
\label{eq:g3}T_{ab}= (\rho+p)u_a u_b +pg_{ab} +q_au_b +q_bu_a +
{\pi}_{ab},
\end{equation}
where $\rho$ is the density of the
fluid, $p$ is the isotropic pressure, $q_a$ is the heat flux
vector and ${\pi}_{ab}$ is the stress tensor. The stress tensor can be expressed as
\begin{equation}
\label{eq:g4}{\pi}_{ab}= (p_r -p_t)\left(n_an_b - \frac{1}{3}
h_{ab}\right),
\end{equation}
where $p_r$ is the radial pressure, $p_t$ is the tangential pressure
and  $\mathbf{n}$ is a unit radial vector given by $n^a
=\frac{1}{B}\delta_1^a$. The isotropic pressure
\begin{equation}
\label{eq:g5}p=\frac{1}{3} (p_r+2p_t)
\end{equation}
relates the radial pressure and the tangential pressure.

For the line element (\ref{eq:g1}) and matter distribution
(\ref{eq:g3}) the coupled Einstein field equations become
\begin{subequations}
\begin{eqnarray}
\label{eq:g6a} \rho &=& \frac{2}{A^2}
\frac{\dot{B}}{B}\frac{\dot{Y}}{Y}+\frac{1}{Y^2}+\frac{1}{A^2}\frac{\dot{Y}^2}{Y^2}\nonumber\\
&&- \frac{1}{B^2}\left(2 \frac{Y''}{Y}+ \frac{{Y'}^2}{Y^2}- 2
\frac{B'}{B}\frac{Y'}{Y}\right),\\
\label{eq:g6b} p_r &=& \frac{1}{A^2} \left(-2 \frac{\ddot{Y}}{Y}
-\frac{\dot{Y}^2}{Y^2} + 2
\frac{\dot{A}}{A}\frac{\dot{Y}}{Y}\right)\nonumber\\
&& +\frac{1}{B^2} \left(\frac{{Y'}^2}{Y^2}
+2\frac{A'}{A}\frac{Y'}{Y}\right)- \frac{1}{Y^2},\\
\label{eq:g6c} p_t &=& -\frac{1}{A^2} \left(\frac{\ddot{B}}{B}
-\frac{\dot{A}}{A}\frac{\dot{B}}{B}+\frac{\dot{B}}{B}\frac{\dot{Y}}{Y}-
\frac{\dot{A}}{A}\frac{\dot{Y}}{Y}+\frac{\ddot{Y}}{Y}\right)\nonumber\\
&& +\frac{1}{B^2} \left(\frac{A''}{A}-\frac{A'}{A}\frac{B'}{B}
+\frac{A'}{A}\frac{Y'}{Y}-\frac{B'}{B}\frac{Y'}{Y}+ \frac{Y''}{Y}
\right),\\
\label{eq:g6d}q&=& -\frac{2}{AB^2}\left(-\frac{\dot{Y}'}{Y}
+\frac{\dot{B}}{B}\frac{Y'}{Y}+\frac{A'}{A}\frac{\dot{Y}}{Y}\right),
\end{eqnarray}
\end{subequations}
where the heat flux $q^a =(0,q,0,0)$ has only the nonvanishing
radial component. A comprehensive treatment of the effects of anisotropy with heat flow
in general relativity was carried out by Herrera \emph{et al} \cite{add1}; the
first study with anisotropy appears to be in the treatment of Lemaitre \cite{add2}.
The system of equations
(\ref{eq:g6a})-(\ref{eq:g6d}) governs the general model when describing  matter distributions with anisotropic pressures in the
presence of heat flux for a spherically symmetric relativistic
stellar object. For this model (\ref{eq:g6a})-(\ref{eq:g6d}) describes the nonlinear gravitational
interaction for a shearing matter distribution which is expanding
and accelerating. From (\ref{eq:g6a})-(\ref{eq:g6d}), we observe
that if forms for the gravitational potentials $A, B$ and $Y$ are
known, then the expressions for the matter variables $\rho, p_r,
p_t$ and $q$ follow immediately. When the radial and tangential pressures are identical then $p_r=p_\perp$ which generates
an additional nonlinear partial differential equation called the condition of pressure isotropy.

The Vaidya exterior spacetime \cite{2} of a radiating star is given by
\begin{equation}
\label{eq:g7} ds^2 = - \left(1-\frac{2m(v)}{R}\right)dv^2 -2dvdR+
R^2 (d{\theta^2}+\sin^2\theta d{\phi}^2),
\end{equation}
where $m(v)$ denotes the mass of the fluid as measured by an
observer at infinity. The line element (\ref{eq:g7}) represents
coherent null radiation. The flow of the radiation is restricted to the radial direction
relative to the hypersurface $\Sigma$, which represents the boundary
of the star. The matching of the metric potentials and extrinsic curvature for
the interior spacetime (\ref{eq:g1}) and the exterior spacetime (\ref{eq:g7}) produces junction conditions on the hypersurface $\Sigma$.
These can be written as
\begin{subequations}
\begin{eqnarray}
\label{eq:g8a} A(R_{\Sigma},t)dt &=&\left(1- \frac{2m}{R_{\Sigma}}+
2 \frac{dR_{\Sigma}}{dv}\right)^{\frac{1}{2}}dv,\\
\label{eq:g8b}Y(R_{\Sigma},t) &=& R_{\Sigma}(v),\\
\label{eq:g8c} m(v)_{\Sigma}&=& \left[\frac{Y}{2}\left(1+
\frac{\dot{Y}^2}{A^2}-\frac{{Y'}^2}{B^2}\right)\right]_{\Sigma},\\
\label{eq:g8d} (p_r)_{\Sigma} &=&(qB)_{\Sigma}.
\end{eqnarray}
\end{subequations}
The junctions conditions (\ref{eq:g8a})-(\ref{eq:g8d}) were first derived by Santos \cite{3} for a
shear-free radiating relativistic star.
It is important to note the nonvanishing of the radial pressure at the boundary $\Sigma$. Thus there is an additional
differential equation (\ref{eq:g8d}) which has to be satisfied
together with the system of Einstein field equations (\ref{eq:g6a})-(\ref{eq:g6d}).
Junction conditions similar to (\ref{eq:g8d}) are important in describing phenomena
which arise in astrophysics. Di Prisco \emph{et al} \cite{16} generated junction conditions relevant to spherical
collapse with dissipation, in the presence of shear, for nonadiabatic charged fluids. Causal thermodynamics,
in the context of the Israel-Stewart theory, was utilised by Herrera \emph{et al} \cite{17} to study viscous
dissipative gravitational collapse in both the streaming out and diffusion approximations.

\section{The boundary condition}
Substituting (\ref{eq:g6b}) and (\ref{eq:g6d}) in (\ref{eq:g8d}) we
obtain the boundary condition which has to be satisfied at the stellar surface.
\begin{eqnarray}
\label{eq:g9}&& 2 Y \ddot{Y}+\dot{Y}^{2}-2
\left(\frac{\dot{A}}{A}+\frac{A'}{B}\right)Y\dot{Y}+2
\frac{A}{B}Y\dot{Y}' \nonumber\\
&& -2 \frac{A}{B^2} \left(A' +\dot{B}\right)Y Y'-
\frac{A^2}{B^2}{Y'}^2 +A^2=0.
\end{eqnarray}
\\
Equation (\ref{eq:g9}) is the governing equation
that determines the gravitational behaviour of the radiating anisotropic star with nonzero shear, acceleration and
expansion. It is clear that (\ref{eq:g9}) is highly nonlinear; it is difficult to solve without making certain
simplifying assumptions. Some exact solutions to (\ref{eq:g9}) were found by Naidu \emph{et al} \cite{11} 
and Rajah and Maharaj \cite{12} for particles in geodesic motion ($\dot u^a=0$) but 
expansion $\Theta \ne 0$ and shear $\sigma \ne 0$. Chan \cite{18} considered the general
case with $\dot u^a\ne 0$, $\Theta \ne 0$ and shear $\sigma \ne 0$ but no exact solutions were found. Instead the boundary condition
was analysed numerically to study the physical features of the model, producing a final state where the star has radiated away mass during collapse. By assuming a relation between the metric functions $B$ and $Y$ for Euclidean stars Govender \emph{et al} \cite{15} found particular models with shear.

Our intention is to solve (\ref{eq:g9}) exactly without restricting the functions. For convenience we rewrite (\ref{eq:g9}) in the following form
\begin{equation}
\label{eq:g10}
\dot{B}-\left[\frac{\ddot{Y}}{AY'}+\frac{\dot{Y}^2}{2AYY'}-
\frac{\dot{A}}{A^2}\frac{\dot{Y}}{Y'}+\frac{A}{2YY'}\right]B^2  -
\left[\frac{\dot{Y}'}{Y'}-\frac{A'}{A}\frac{\dot{Y}}{Y'}\right]B
+\left[A' + \frac{AY'}{2Y}\right]=0.
\end{equation}
In general (\ref{eq:g10}) is a Riccati equation in the gravitational potential $B$. This Riccati equation can be solved in special cases.

\section{Exact solutions}
The complexity and nonlinearity in (\ref{eq:g10}) makes it difficult to solve in general. However particular exact solutions can be found if
we view (\ref{eq:g10}) as a first order differential equation in the variable $B$ and place restrictions on the bracketed expressions.
We demonstrate this in the following three cases.

\subsection{Linear equation}
Note that equation (\ref{eq:g10}) becomes a linear equation if we
set
\begin{equation}
\frac{\ddot{Y}}{AY'}+\frac{\dot{Y}^2}{2AYY'}-
\frac{\dot{A}}{A^2}\frac{\dot{Y}}{Y'}+\frac{A}{2YY'}=0.
\end{equation}
This equation can be written as
\begin{equation}
\dot{A}- \left[\frac{\ddot{Y}}{\dot{Y}}+\frac{\dot{Y}}{2Y}\right]A
= \frac{A^3}{2Y \dot{Y}} ,
\end{equation}
which is a Bernoulli equation in the variable $A$. Even though $Y$ is an arbitrary function, this equation
can be integrated in general, and we have
\begin{equation}
\label{eq:g11} A^2=\frac{Y \dot{Y}^2}{h(r)-Y},
\end{equation}
where $h(r)$ is a function of integration. With the result (\ref{eq:g11}), we find that (\ref{eq:g10}) becomes
\begin{equation}
\label{eq:g12} \dot{B} -
\left[\frac{\dot{Y}'}{Y'}-\frac{A'}{A}\frac{\dot{Y}}{Y'}\right]B
+\left[A' + \frac{AY'}{2Y}\right]=0,
\end{equation}
which is linear in $B$.

The bracketed expressions in (\ref{eq:g12}) contain the functions $A, Y$ and their
 derivatives. In spite of this difficulty it is possible to solve (\ref{eq:g12}) and obtain $B$ in general. Therefore the solution for the junction condition
(\ref{eq:g10}) can be given by
\begin{subequations}
\label{eq:g13}
\begin{eqnarray}
\label{eq:g13a}A&=& \sqrt{\frac{Y \dot{Y}^2}{h(r)-Y}},\\
\label{eq:g13b} B&=& Y' \exp \left(- \int \frac{A'\dot{Y}}{A
Y'}dt\right) \times \nonumber \\
&& \left\{k(r)-\displaystyle{\int} \left[
\left(\frac{A'}{Y'}+\frac{A}{2Y}\right) \exp \left(\int
\frac{A'\dot{Y}}{A Y'}dt\right) \right] dt \right \}, \\
\label{eq:g13c}Y &=& Y(t,r),
\end{eqnarray}
\end{subequations}
where $k(r)$ is a function of integration. We believe that
(\ref{eq:g13a})-(\ref{eq:g13c}) is a new solution to the boundary
condition (\ref{eq:g9}). Note that the gravitational potential
$Y(t,r)$ is an arbitrary function in this class of solution. Once
$Y$ is specified then an explicit form for $A$ is generated from
(\ref{eq:g11}) and the integrals in
(\ref{eq:g13a})-(\ref{eq:g13c}) can be evaluated. Consequently
explicit forms for the metric functions $A, B$ and $Y$ can be
found. The choice for $Y$ should be made to provide a physically
reasonable model.

\subsection{Bernoulli equation}
Observe that equation (\ref{eq:g10}) reduces to a Bernoulli
equation if we set
\begin{equation}
A' + \frac{AY'}{2Y}=0.
\end{equation}
Integrating this equation we get
\begin{equation}
\label{eq:g14}Y =\frac{C_1(t)}{A^2},
\end{equation}
where  $C_1(t)$ is a function of integration. Substituting
(\ref{eq:g14}) into (\ref{eq:g10}) we obtain
\begin{eqnarray}
\label{eq:g15}&& \dot{B}-\left[\frac{3}{2}\frac{\dot{C_1}}{C_1}- 4
\frac{\dot{A}}{A}+\frac{\dot{A}'}{A'}\right]B \nonumber \\
&&= \left[\frac{7}{2}\frac{\dot{A}\dot{C_1}}{A A'C_1}-5
\frac{\dot{A}^2}{A^2A'}-\frac{\ddot{C_1}}{2C_1A'}+\frac{\ddot{A}}{AA'}
-\frac{\dot{C_1}^2}{4C_1^2A'}-\frac{A^6}{4C_1^2A'} \right]B^2,
\end{eqnarray}
which is a Bernoulli equation in the variable $B$.

The coefficients in (\ref{eq:g15}) contain the functions $A$,
$C_1$ and their derivatives; however it can be integrated in
general. On integrating (\ref{eq:g15}) we can write
\begin{equation}
\label{eq:g16}B=\frac{A' C_1^{3/2}}{A^4 [ \int I dt +g(r)]},
\end{equation}
where $g(r)$ is a function of integration and for convenience we
have defined 
\begin{equation}I= -\frac{7}{2}\frac{C_1^{1/2}
\dot{A} \dot{C_1}}{A^5}+5 \frac{\dot{A}^2 C_1^{3/2}}{A^6}
+\frac{\ddot{C_1} C_1^{1/2}}{2A^4}
-\frac{\ddot{A}C_1^{3/2}}{A^5}+\frac{\dot{C_1}^2}{4C_1^{1/2}A^4}+\frac{A^2}{4C_1^{1/2}}.
\end{equation}
Therefore the functions
\begin{subequations}
\label{eq:g17}
\begin{eqnarray}
\label{eq:g17a}&& A= A(t,r),\\
\label{eq:g17b}&& B=\frac{A' C_1^{3/2}}{A^4 [ \int I dt +g(r)]},\\
\label{eq:g17c}&&Y =\frac{C_1}{A^2},
\end{eqnarray}
\end{subequations}
satisfy the junction condition (\ref{eq:g10}). The model (\ref{eq:g17}) is an exact solution to the boundary condition (\ref{eq:g9}).
Note that the gravitational potential $A(t,r)$ is an arbitrary function in this class of solution. Once $A$ is specified,
together with the integration constants $C_1$, then an explicit form for $I$ can be determined. Then the metric functions
 $A, B$ and $Y$ can be expressed in closed form in terms of elementary or special functions. The choice for $A$ should be made on physical grounds.

\subsection{Inhomogeneous Riccati equation}
Note that equation (\ref{eq:g10}) has the form of an inhomogeneous
Riccati equation if we set
\begin{equation}\frac{\dot{Y}'}{Y'}-\frac{A'}{A}\frac{\dot{Y}}{Y'}=0.
\end{equation}
Integrating this equation we get
\begin{equation} A= \dot{Y}
\alpha(t), 
\end{equation}
where $\alpha(t)$ is a function of  integration. In this
case (\ref{eq:g10}) becomes
\begin{equation}
\label{eq:g18} \dot{B}=
\left[\frac{\dot{Y}(1+\alpha^2)}{2 \alpha YY'}-\frac{\dot{\alpha}}{\alpha^2Y'}\right]B^2
-\left[\dot{Y}'\alpha +\frac{\dot{Y}Y'\alpha}{2Y} \right].
\end{equation}
This is an inhomogenous Riccati equation which is difficult to
analyse in general. However we shall show that it is possible to
integrate this equation by placing restrictions on the functions
$\alpha$ and $Y$.

If we take  $\alpha$  to be a real constant and $Y$ to be the separable
function
\begin{equation}
\label{eq:g19} Y(t,r)=K(r)C(t),
\end{equation}
where  $K(r)$ and $C(t)$ are arbitrary functions of $r$ and $t$
respectively, then equation (\ref{eq:g18}) becomes
\begin{equation}
\label{eq:g20}  \dot{B}= \frac{(1+{\alpha}^2)}{2 \alpha
K'}\frac{\dot{C}}{C^2} B^2 -\frac{3}{2} \alpha K' \dot{C}.
\end{equation}
The Riccati equation (\ref{eq:g20}) is not in standard form.
Consequently we introduce the transformation
\begin{equation}
B =w C 
\end{equation}
to obtain
\begin{equation}
\label{eq:g21}
\left[\frac{2\alpha K'}{(1+\alpha^2)w^2-2\alpha K'w-3\alpha^2K'^2}\right]\dot w=\frac{\dot C}{C}.
\end{equation}
The advantage of the form (\ref{eq:g21}) is that it is a separable
equation in the variables $w$ and $C$. Equation (\ref{eq:g21}) can
be integrated if the constant $\alpha $ is specified. To
demonstrate a simple exact solution we take $\alpha =-2$. Then
(\ref{eq:g21}) can be written as
\begin{equation}
\frac{\dot{w}}{(5w-6K')(w+2K')}= - \frac{1}{4
K'}\frac{\dot{C}}{C}.
\end{equation}
On integrating the above equation we obtain
\begin{equation}
w=\frac{2 K' [3C^4+f(r)]}{5 C^4-f(r)},
\end{equation}
where $f(r)$ is a function of integration. The metric function $B$
then follows since $B=wC$.

Therefore we have generated a new solution to the inhomogenous Riccati equation (\ref{eq:g18}). The form of the solution is given by
\begin{subequations}
\label{eq:g22}
\begin{eqnarray}
\label{eq:g22a} A &=& -2K \dot{C},\\
\label{eq:g22b} B &=&\frac{2 K' C[3 C^4+f(r)]}{5 C^4-f(r)},\\
\label{eq:g22c}Y &=& KC.
\end{eqnarray}
\end{subequations}
The form of the solution (\ref{eq:g22a})-(\ref{eq:g22c}) is particularly simple and does not involve further integration. The functions
$C, K$ and $f$ are arbitrary; the physics of a specific model investigated will determine their explicit form. It is remarkable that the explicit solution (\ref{eq:g22a})-(\ref{eq:g22c}) can be found for the boundary condition (\ref{eq:g9}) in this case; Riccati equations are difficult to solve
and only limited classes of solution are known to exist.

\section{Example}
The simple forms of the solutions found this paper make it possible to study the physical behaviour of the model.
In this section we briefly consider the
physical features of the solution generated in Section 4.3. For the
gravitational potentials obtained in
(\ref{eq:g22a})-(\ref{eq:g22c}), we take $C(t)=t^2, K(r)=r$ and $f(r)=k$, where $k$ is a real constant. For these values the
kinematical quantities become
\begin{subequations}
\begin{eqnarray}
\label{eq:g24a} \dot{u}^a &=& \left(0,
\frac{[5 {\tilde t}-1]^2}{4 r [3\tilde t+1]^2 \sqrt{k\tilde t}},0 ,0\right),\\
\label{eq:g24b} \Theta &=& \frac{[3 +26 \tilde t -45\tilde
t^{2}]}{2r[-1 +2\tilde t+15\tilde t^{2}](k\tilde t)^\frac{1}{4}},\\
\label{eq:g24c} \sigma &=&\frac{16 \tilde t}{3r [1
-2\tilde t-15\tilde t^{2}](k\tilde t)^\frac{1}{4}},
\end{eqnarray}
\end{subequations}
where we have set $\tilde t=t^8/k$ for convenience. From (\ref{eq:g24a})-(\ref{eq:g24c}) we observe that the
acceleration $\dot{u}^a$, the expansion $\Theta$ and the magnitude
of the shear scalar are nonzero. These quantities remain finite in the interior apart from the stellar centre. Also note that in the limiting case as $t \rightarrow \infty$
the acceleration $\dot{u}^a \rightarrow 0$ the shear scalar
$\sigma \rightarrow 0$ and expansion $\Theta \rightarrow 0$. As the model evolves for large time the kinematical quantities
grow progressively smaller. From the forms of $\dot u^a$ and $\Theta$ given above we observe that the acceleration decreases more rapidly
than the expansion for large time.

The matter variables become
\begin{subequations}
\begin{eqnarray}
\label{eq:g25a} \rho &=& \frac{\left[-3 -43\tilde t
+15\tilde t^{2}+95\tilde t^{3}\right]}{2r^2 [3\tilde t+1]^2
[5\tilde t-1]\sqrt{k\tilde t}},\\
\label{eq:g25b} p_r &=& \frac{1}{4r^2
\sqrt{k\tilde t}}\left[\frac{3[5\tilde t-1]^2}{[3\tilde t+1]^2}-5\right],\\
\label{eq:g25c}p_t &=& \frac{4[-\frac{13}{k} -45\tilde t-95\tilde
t^{2}+25\tilde t^{3}]}
{r^2k[1 -2\tilde t-15\tilde t^{2}]^2},\\
\label{eq:g25d} q&=&
\frac{[1+25\tilde t-165\tilde t^{2}+75\tilde t^{3}]}{4r^2[3\tilde t+1]^3(k\tilde t)^\frac{3}{4}}.
\end{eqnarray}
\end{subequations}
From (\ref{eq:g25a})-(\ref{eq:g25d}) we observe that the energy
density $\rho$, radial pressure $p_r$, tangential pressure $p_t$ and
heat flux $q$ are continuous in the stellar interior, apart from the centre. At later times as $t \rightarrow \infty $ we note that
$q \rightarrow 0$ so that the heat flux is radiated away during the process of gravitational collapse. It is interesting to see that the energy density
$\rho$, radial pressure $p_r$, the tangential pressure $p_t$ and the heat flux $q$ are proportional to $r^{-2}$, and are decreasing functions
as we approach the boundary of the star. The behaviour that $\rho \propto r^{-2}$ is of physical importance. It is interesting to observe that this
property is also present in Newtonian isothermal spheres and relativistic isothermal cosmological models as pointed out by Saslaw \emph {et al} \cite{19}.

\begin{figure}
\centering
\includegraphics[angle = 0,scale = 1.2 ]{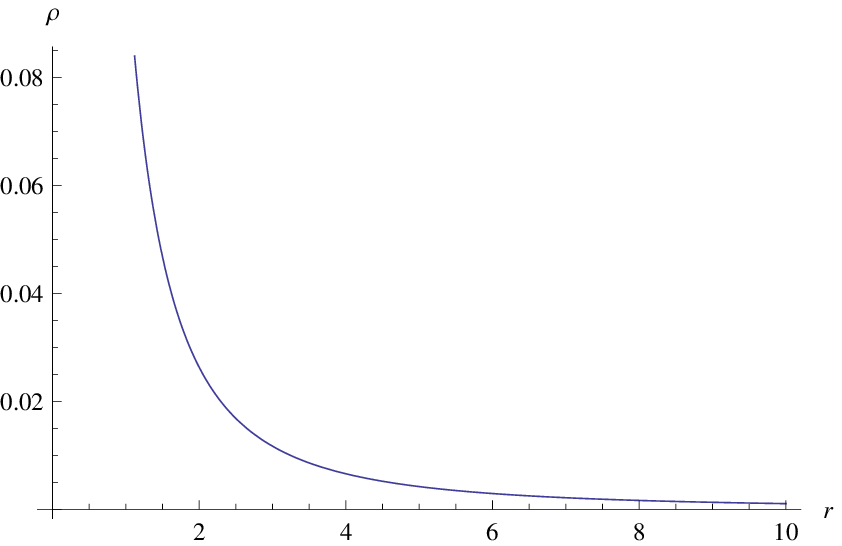}
\caption{\label{fig1} Density}
\end{figure}

\begin{figure}
\centering
\includegraphics[angle = 0,scale = 1.2 ]{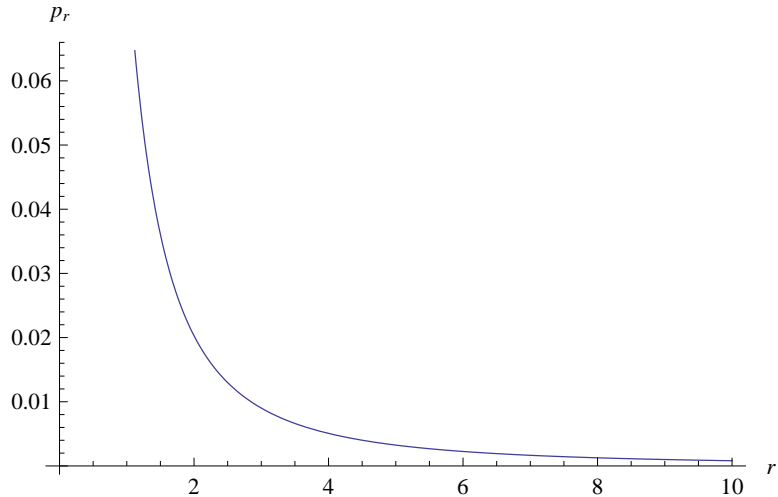}
\caption{\label{fig2} Radial pressure}
\end{figure}

\begin{figure}
\centering
\includegraphics[angle = 0,scale = 1.2 ]{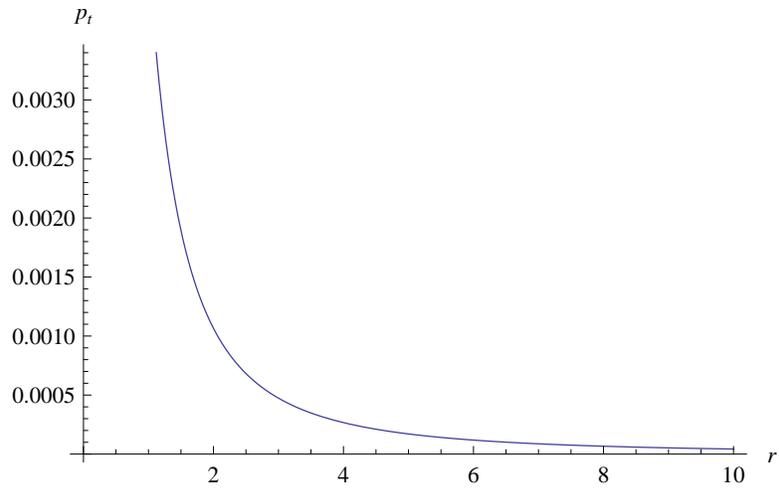}
\caption{\label{fig3} Tangential pressure}
\end{figure}

\begin{figure}
\centering
\includegraphics[angle = 0,scale = 1.2 ]{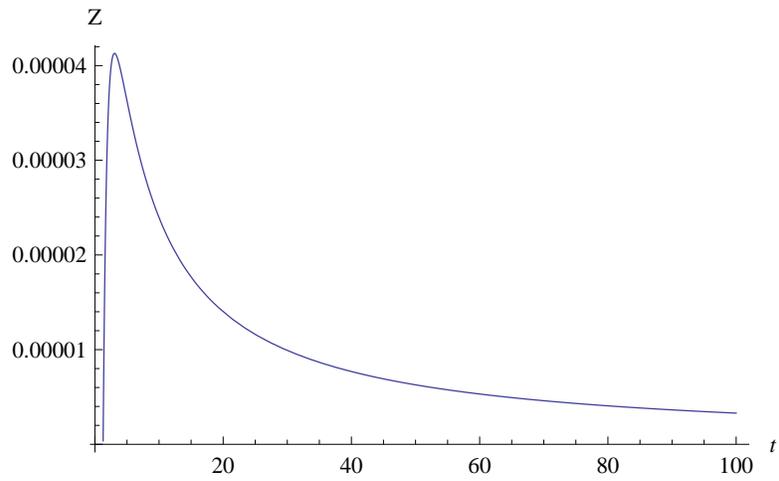}
\caption{\label{fig4} $Z=(\rho+p_r)^2-4q^2$}
\end{figure}

\begin{figure}
\centering
\includegraphics[angle = 0,scale = 1.2 ]{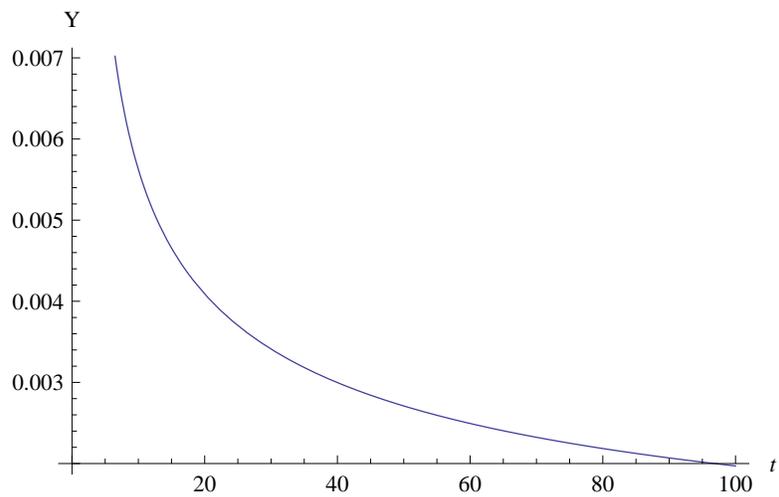}
\caption{\label{fig5} $Y=\rho-p_r-2p_t+[(\rho+p_r)^2-4q^2]^\frac{1}{2}$}
\end{figure}

\begin{figure}
\centering
\includegraphics[angle = 0,scale = 1.2 ]{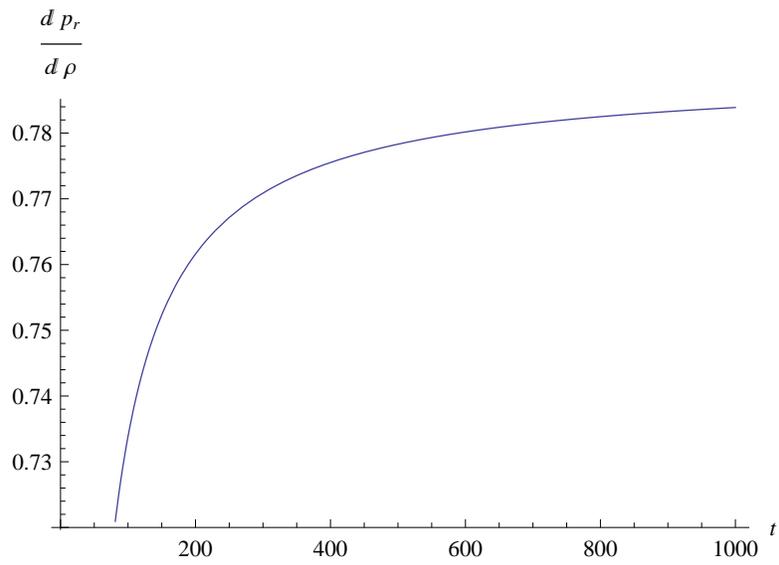}
\caption{\label{fig6} Sound speed}
\label{fig:abc}
\end{figure}

A qualitative analysis of the matter variables, energy conditions, and stability is difficult to achieve for the interior matter distribution.
 However it is possible to generate graphical plots which indicate physical viability. In Fig. \ref{fig1}-\ref{fig3} we have 
 plotted the energy density $\rho$, the radial pressure $p_r$ and the tangential pressure $p_t$. 
 We observe that $\rho>0, p_r>0$ and $p_t>0$. In addition we have the behaviour $\rho' <0$ and $p_r'<0$
  so that $\rho$ and $p$ are decreasing functions outwards from the centre to the stellar 
  surface. For fixed values of the radial coordinate  it is possible to plot the behaviour of
\begin{eqnarray}
Z &=& (\rho+p_r)^2-4q^2\\
Y &=& \rho-p_r-2p_t+[(\rho+p_r)^2-4q^2]^\frac{1}{2}
\end{eqnarray}
Typical behaviour of these quantities are represented in Fig. \ref{fig4} and Fig. \ref{fig5} respectively. 
These graphs show that $Z>0$ and $Y>0$. The behaviour exhibited in this physical analysis
 indicates that the weak, strong and dominant conditions are satisfied in interior points 
 away from the centre. Also note from Fig. \ref{fig6}  that the speed of sound is less than the speed of light so that causality is not violated.

\begin{figure}
\centering
\includegraphics[angle = 0,scale = 1.2 ]{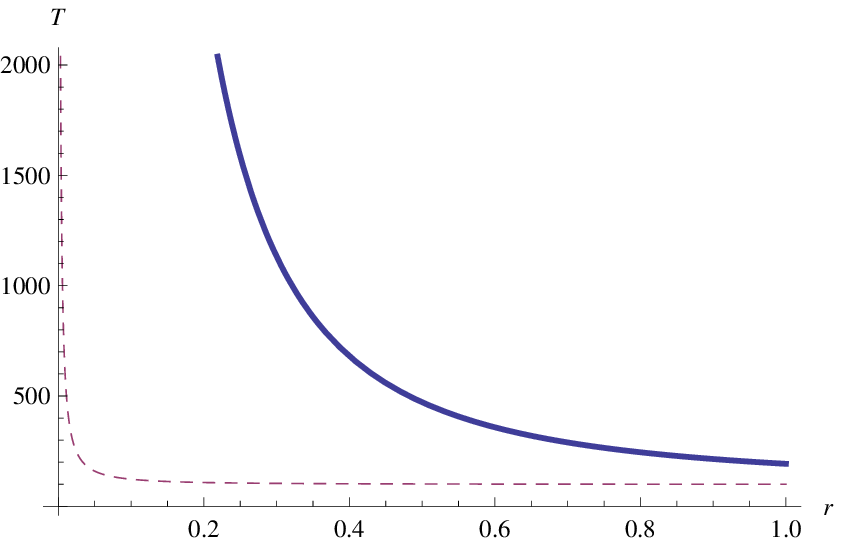}
\caption{\label{fig7} Temperature: Acausal(Dashed); Causal(Solid)}
\end{figure}

Next we briefly consider the relativistic effect of the causal
temperature in our model. The Maxwell-Cattaneo heat transport
equation, in the absence of rotation and viscous stresses is given by
\begin{equation}
\label{eq:g26} \tau h_a^{~b}\dot{q}_b+q_a = -\kappa
\left(h_a^{~b}{\nabla}_b T+T \dot{u}_a\right),
\end{equation}
where $\tau$ is the relaxation time, $\kappa$ is the thermal
conductivity, $h_{ab}=g_{ab}+u_au_b$ projects into the comoving
rest space and $T$ is the local causal temperature. Equation (\ref{eq:g26}) reduces to the acausal
Fourier heat transport equation when $\tau =0$. The causal transport equation (\ref{eq:g26}) can be
written as
\begin{equation}
\label{eq:g27} T(t,r)= -\frac{1}{\kappa A} \int \left[\tau
\dot{(qB)}B+ AqB^2 \right]dr
\end{equation}
for the metric (\ref{eq:g1}).
Martinez \cite{20}, Govender \emph{et al} \cite{21} and Di Prisco \emph{et
al} \cite{22} have shown that the relaxation time $\tau$ has a major effect on the thermal evolution, particularly in the latter stages of
collapse. Rajah and Maharaj \cite{12} and Naidu \emph{et al} \cite{11}
showed that in the presence of shear stress, the relaxation time
decreases as the collapse proceeds and the central temperature
increases. For our case, (\ref{eq:g27}) becomes
\begin{eqnarray}
\label{eq:g28} T(\tilde t,r)&=& \frac{\tau [-1 +15 \tilde t -315\tilde 
t^{2}+45 \tilde t^{3}]}{\kappa r^2[3\tilde t
+1]^2[5\tilde t-1]\sqrt{k\tilde t}} \nonumber\\
&&+\frac{[1 +30\tilde t-15\tilde t^{2}(k\tilde t)^\frac{1}{4}]\ln[r]}{\kappa
r[-1 +2\tilde t+15 \tilde t^{2}](k\tilde t)^\frac{1}{4}}+h(t),
\end{eqnarray}
where $h(t)$ is a function of integration and we set $\tau$ and
$\kappa$ as constant. When $\tau =0$, we can regain the acausal
(Eckart) temperature from (\ref{eq:g28}). It is possible to plot the causal and acausal temperatures against the
 radial coordinate. In Fig. \ref{fig7} the temperature profiles are similar to the 
 curves in Rajah and Maharaj\cite{12}. The temperature is a decreasing function from the centre to the boundary of the star in
both the causal and acausal curves. The inclusion of particle acceleration
in our model may contribute to the more rapid decrease of temperature from the core to the boundary of the star. This may be applicable
to phases of collapse where there is rapid expansion and cooling of the outer layers of the stellar fluid. As in the Rajah and Maharaj \cite{12} model it is clear that the causal
temperature is greater than the acausal temperature throughout the stellar interior.

\section{Discussion}
In summary, we considered the general case of a spherically
symmetric radiative star undergoing gravitational collapse when the interior spacetime consists of an
accelerating, expanding and shearing matter distribution. The junction condition is rewritten so that it can be considered
as a first order equation in the potential $B(t,r)$. It is then possible to consider the junction condition as a standard
differential equation: a linear equation, a Bernoulli equation and a Riccati equation. The linear and Bernoulli equations are solved in general.
The Riccati equation can only be solved for a particular value of the integration constant. Therefore three new classes of solutions to the boundary
condition have been found. For a particular metric, corresponding to the inhomogenous Riccati equation, it is possible to obtain forms for the
kinematical quantities and matter variables. It is then possible to indicate the behaviour of the model for large values of time.

\section*{Acknowledgements}
ST and SSR thank the National Research Foundation and the University of
KwaZulu-Natal for financial support. ST is grateful to Eastern
University, Sri Lanka, for study leave. SDM acknowledges that this
work is based upon research supported by the South African Research
Chair Initiative of the Department of Science and Technology and the
National Research Foundation. We are grateful to the referee for comments
that have substantially improved the manuscript.

\end{document}